\documentclass{aa}
\usepackage{graphicx}
\usepackage{bm}
\usepackage{times}
\usepackage{amsmath}
\usepackage{amssymb}
\usepackage{natbib}
\usepackage{color}
\usepackage{mathrsfs}
\usepackage[colorlinks,allcolors=blue]{hyperref}
\usepackage{url}
\usepackage{multirow}
\bibpunct{(}{)}{;}{a}{}{,}
\graphicspath{{./fig/}{./png/}}
\newcommand{\DD}{{\rm D}}

\newcommand{\bbb}{{\bm b}}

\newcommand{\uuu}{{\bm u}}

\newcommand{\ooo}{{\bm \omega}}

\newcommand{\BBB}{{\bm B}}
\newcommand{\mBBB}{\overline{\bm B}}

\newcommand{\UUU}{{\bm U}}
\newcommand{\mUUU}{\overline{\bm U}}

\newcommand{\mUUi}{\overline{U}_i}
\newcommand{\mUUj}{\overline{U}_j}

\newcommand{\mUUr}{\overline{U}_r}

\newcommand{\mUUt}{\overline{U}_\theta}
\newcommand{\mUUp}{\overline{U}_\phi}

\newcommand{\mBBi}{\overline{B}_i}
\newcommand{\mBBj}{\overline{B}_j}

\newcommand{\mBBp}{\overline{B}_\phi}

\newcommand{\Eq}[1]{Eq.~(\ref{#1})}

\newcommand{\Equsa}[2]{Eqs.~(\ref{#1}) to (\ref{#2})}
\newcommand{\EQ}{\begin{equation}}
\newcommand{\EN}{\end{equation}}
\newcommand{\EQA}{\begin{eqnarray}}
\newcommand{\ENA}{\end{eqnarray}}
\newcommand{\brac}[1]{\langle #1 \rangle}
\newcommand{\pd}{\partial}

\newcommand{\mean}[1]{\overline{#1}}

\newcommand{\cP}{c_{\rm P}}
\newcommand{\cV}{c_{\rm V}}

\newcommand{\urms}{u_{\rm rms}}

\newcommand{\orms}{\omega_{\rm rms}}

\newcommand{\chiSGS}{\chi_{\rm SGS}}

\newcommand{\Co}{{\rm Co}}

\newcommand{\Pe}{{\rm Pe}}
\newcommand{\Pra}{{\rm Pr}}

\newcommand{\PraSGS}{{\rm Pr}_{\rm SGS}}

\newcommand{\PrM}{{\rm Pr}_{\rm M}}
\newcommand{\Pm}{{\rm Pm}}

\newcommand{\Ra}{{\rm Ra}}

\newcommand{\Rey}{{\rm Re}}

\newcommand{\ReM}{{\rm Re}_{\rm M}}

\newcommand{\Ta}{{\rm Ta}}

\newcommand{\mOm}{\mean{\Omega}}

{}

\newcommand{\rin}{r_{\rm in}}
\newcommand{\Fbot}{F_{\rm bot}}

\def\onethird{{\textstyle{1\over3}}}

\def\onehalf{{\textstyle{1\over2}}}

\def\threehalfs{{\textstyle{3\over2}}}

\newcommand{\Figa}[1]{Fig.~\ref{#1}}

\newcommand{\Figus}[2]{Figures~\ref{#1} and \ref{#2}}
\newcommand{\Figust}[2]{Figures~\ref{#1} to \ref{#2}}
\newcommand{\Figu}[1]{Figure~\ref{#1}}

\newcommand{\Seca}[1]{Sect.~\ref{#1}}
\newcommand{\Table}[1]{Table~\ref{#1}}

\begin{document}

\authorrunning{K\"apyl\"a}
\titlerunning{Transition from anti-solar to solar-like differential rotation}

   \title{Transition from anti-solar to solar-like differential
     rotation: Dependence on Prandtl number}

   \author{P. J. K\"apyl\"a
          \inst{1,2}
          }

   \institute{Georg-August-Universit\"at G\"ottingen, Institut f\"ur 
              Astrophysik und Geophysik, Friedrich-Hund-Platz 1, D-37077 G\"ottingen,
              Germany
              email: \href{mailto:pkaepyl@uni-goettingen.de}{pkaepyl@uni-goettingen.de}
   \and
              Nordita, KTH Royal Institute of Technology and Stockholm
              University, Stockholm, Sweden
}

\date{\today}

\abstract{Late-type stars such as the Sun rotate differentially due to
  the interaction of turbulent convection and rotation.
   }%
   {The aim of the study is to investigate the effects of the
     effective thermal Prandtl number, which is the ratio of kinematic
     viscosity to thermal diffusivity, on the transition from
     anti-solar (slow equator, fast poles) to solar-like (fast
     equator, slow poles) differential rotation.
   }%
   {Three-dimensional hydrodynamic and magnetohydrodynamic simulations
     in semi-global spherical wedge geometry are used to model
     convection zones of solar-like stars.
   }%
   {The overall convective velocity amplitude increases as the Prandtl
     number decreases in accordance with earlier studies. The
     transition from anti-solar to solar-like differential rotation is
     insensitive to the Prandtl number for Prandtl numbers below unity
     but for Prandtl numbers greater than unity, solar-like
     differential rotation becomes significantly harder to
     excite. Magnetic fields and more turbulent regimes with higher
     fluid and magnetic Reynolds numbers help in achieving solar-like
     differential rotation in near-transition cases where anti-solar
     rotation is found in more laminar simulations. Solar-like
     differential rotation occurs only in cases with radially outward
     angular momentum transport at the equator. The dominant
     contribution to such outward transport near the equator is due to
     prograde propagating thermal Rossby waves.
   }%
   {The differential rotation is sensitive to the Prandtl number only
     for large Prandtl numbers in the parameter regime explored in the
     current study. Magnetic fields have a greater effect on the
     differential rotation, although the inferred presence of a
     small-scale dynamo does not lead to drastically different results
     in the present study. The dominance of the thermal Rossby waves
     in the simulations is puzzling given the non-detection in the
     Sun. The current simulations are shown to be incompatible with
     the currently prevailing mean-field theory of differential
     rotation.
   }%
   \keywords{   turbulence -- convection
   }

  \maketitle


\section{Introduction}

The interplay of turbulent convection with the overall rotation of the
Sun is the primary cause of differential rotation observed at the
solar surface and in the interior
\citep[e.g.][]{R89,MT09}. Three-dimensional numerical simulations
solving the equations of magnetohydrodynamics (MHD) capture the
essence of this process and routinely produce solutions that are
qualitatively similar to the Sun with equatorial acceleration
\citep[e.g.][]{Gi83,BMT04,GSKM13,KKB14}. However, it has become
increasingly clear recently that even the most sophisticated current
simulations are missing something essential. The most striking
manifestation of this is that simulations using nominal solar
luminosity and rotation rate often produce anti-solar (AS)
differential rotation with equatorial deceleration
\citep[e.g.][]{FF14,KKB14,HRY15a}, whereas solar-like (SL)
differential rotation is achieved only with significantly more rapid
rotation \citep[e.g.][]{2018A&A...616A.160V,2020ApJ...898..111M}.

This is related to the convective conundrum
\citep{2016AdSpR..58.1475O} which is essentially the tension between
large-scale velocity amplitudes in simulations in comparison to the
Sun
\citep[e.g.][]{HDS12,2016AnRFM..48..191H,2020RvMP...92d1001S}. Until
recently, the most common way to ensure SL differential rotation in
simulations with solar luminosity and rotation rate has been to lower
the convective velocities by artificially enhancing the radiative
diffusivity \citep[e.g.][]{FF14,KKB14,HRY16}. This, however, cannot be
justified based on physical grounds since convection is thought to
carry practically all of the energy flux through the solar convection
zone (CZ) with the exception of very deep layers. Another, more
plausible, effect is due to magnetic fields: is is conceivable that
sufficiently strong fields can suppress convection to a degree where
the differential rotation flips from AS to SL. Early results with
relatively low resolution simulations were mixed: \cite{KKKBOP15}
found essentially no dependence on magnetic field while \cite{FF14}
and \cite{2015ApJ...810...80S} reported more positive
outcomes. Nevertheless, these simulations most probably did not have
high enough magnetic Reynolds numbers to excite a small-scale
dynamo. This was addressed by recent high-resolution simulations of
\cite{2021NatAs...5.1100H} and \cite{2022arXiv220204183H} which
suggest that SL differential rotation can indeed be achieved with the
help of an efficient small-scale dynamo.

Another important parameter is the Prandtl number, $\Pra = \nu/\chi$,
where $\nu$ is the kinematic viscosity and $\chi$ is the thermal
diffusivity. A notion that the solar convection zone is operating in a
regime where the effective Prandtl number is large, has gained
popularity recently
\citep[e.g.][]{2016AdSpR..58.1475O,2017ApJ...851...74B,2018PhFl...30d6602K}. While
these studies indicate that the overall velocity amplitudes are
decreased in such set-ups, the problem with the differential rotation
becomes actually worse \citep{2018PhFl...30d6602K}. This is because it
is not only the velocity amplitude that is sensitive to $\Pra$, but
also turbulent transport of angular momentum and heat are affected
\citep[e.g.][]{CBTMH91,2021A&A...655A..78K}. Furthermore, theoretical
arguments suggest that $\Pra \ll 1$ in the solar CZ
\citep[e.g.][]{O03,2020RvMP...92d1001S}.

Prandtl numbers deviating strongly from unity are challenging
numerically and therefore most simulations are done in the $\Pra
\approx 1$ regime. It is commonly acknowledged that reaching realistic
parameter regimes in terms of, for example, Prandtl, Reynolds and
Rayleigh numbers with current or foreseeable simulations of stellar
convection is infeasible \citep[e.g.][]{2017LRCA....3....1K}. The main
aim of the present study is to vary the Prandtl number within the
range that is reasonably realizable with numerical simulations with
values above and below unity. The current study is also inspired by
recent results from hydrodynamic non-rotating convection in Cartesian
geometry \citep{2021A&A...655A..78K}, where the convective energy
transport and velocity statistics were found to be sensitive to the
effective Prandtl number.

\section{The model} \label{sect:model}

The simulation set-up is similar to those used in
\cite{2019GApFD.113..149K} and \cite{2020GApFD.114....8K}. The
simulation domain is a spherical wedge that spans $\rin<r<R$ in
radius, where $\rin=0.7R$ and $R$ is the radius of the star,
$\theta_0<\theta<\pi-\theta_0$ in colatitude, where $\theta_0=\pi/12$,
and $0<\phi<\pi/2$ in longitude. Equations of fully compressible MHD
are solved
\begin{eqnarray}
\frac{\pd {\bm A}}{\pd t} & = & {\bm U} \times {\bm B} - \eta \mu_0 {\bm J}, \label{equ:indu}\\
\frac{\DD \ln\rho}{\DD t} & = & -\bm\nabla\bm\cdot{\bm U}, \label{equ:conti}\\
\frac{\DD {\bm U}}{\DD t} & = &  {\bm g}  - 2\bm\Omega_0\times{\bm U} - \frac{1}{\rho}(\bm\nabla p - {\bm J} \times {\bm B} - \bm\nabla \bm\cdot 2\nu\rho\bm{\mathsf{S}}),\label{equ:NS}\\
T\frac{\DD s}{\DD t} & = & \frac{1}{\rho} \left[ \eta\mu_0 {\bm J}^2 - \bm\nabla\bm\cdot( {\bm F}^{\rm rad} + {\bm F}^{\rm SGS} )\right] + 2\nu \bm{\mathsf{S}}^2, \label{equ:ss}
\end{eqnarray}
where ${\bm A}$ is the magnetic vector potential, ${\bm U}$ is the
velocity, ${\bm B} = \bm\nabla\times{\bm A}$ is the magnetic field,
$\eta$ is the magnetic diffusivity, $\mu_0$ is the permeability of
vacuum, ${\bm J}=\bm\nabla\times{\bm B}/\mu_0$ is the current density,
$\DD/\DD t = \pd/\pd t + {\bm U}\bm\cdot\bm\nabla$ is the advective
time derivative, $\rho$ is the density, ${\bm g}=-\bm\nabla \phi$ is
the acceleration due to gravity, where $\phi = -GM/r$ is a fixed
spherically symmetric gravitational potential, with $G$ and $M$ being
the universal gravitational constant and the stellar mass,
respectively. $\bm\Omega_0=(\cos\theta,-\sin\theta,0)\Omega_0$ is the
angular velocity vector, where $\Omega_0$ is the rotation rate of the
frame of reference, $p$ is the pressure, $\nu$ is the kinematic
viscosity, $T$ is the temperature, and $s$ is the specific entropy
with $Ds=\cV D\ln p-\cP D\ln\rho$, where $\cV$ and $\cP$ are the
specific heat capacities in constant volume and pressure,
respectively. The gas is assumed to obey the ideal gas law,
$p=\mathcal{R}\rho T$, where $\mathcal{R}=\cP-\cV$ is the gas
constant. The rate of strain tensor is given by
\begin{eqnarray}
\mathsf{S}_{ij}\, = \,\onehalf (U_{i;j} + U_{j;i}) - \onethird \delta_{ij} \bm\nabla\bm\cdot {\bm U},
\end{eqnarray}
where the semicolons refer to covariant derivatives
\citep{MTBM09}. The radiative flux is given by
\begin{eqnarray}
{\bm F}^{\rm rad} \,=\, -\,K\bm\nabla T,
\label{equ:Frad}
\end{eqnarray}
where $K$ is the heat conductivity. The latter consists of two parts,
$K = K_1 + K_2$, where $K_1 = K_1(r)$ is a fixed function of height
and $K_2 = K_2(\rho,T)$ is density- and temperature-dependent
according to Kramers opacity law \citep{WHTR04}. The profile of $K_1$
is given by
\begin{eqnarray}
K_1 = K_{\rm top} \left[1 + \tanh\left(\frac{r-R}{d_K}\right)\right],
\end{eqnarray}
where $K_{\rm top}={2 \over 3} \Fbot$, with $\Fbot=L/4\pi\rin^2$ where
$L$ is the luminosity of the star, and where $d_K = 0.015R$. The
contribution $K_2$ is given by
\begin{eqnarray}
K_2(\rho,T)\, =\, K_0 (\rho/\rho_0)^{-(a+1)} (T/T_0)^{3-b},
\label{equ:Krad2}
\end{eqnarray}
where $\rho_0$ and $T_0$ are reference values of density and
temperature, and the values $a=1$ and $b=-7/2$ correspond to the
Kramers opacity law. This formulation was first used in convection
simulations by \cite{2000gac..conf...85B}.

The subgrid scale (SGS) flux is given by
\begin{eqnarray}
  {\bm F}^{\rm SGS}\, = \,-\,\chiSGS \rho \bm\nabla s',
\label{equ:FSGS}
\end{eqnarray}
where $\chiSGS$ is the (constant) SGS diffusion coefficient for the
entropy fluctuation $s'(r,\theta,\phi)=s-\brac{s}_{\theta\phi}$, where
$\brac{s}_{\theta\phi}$ is the spherically symmetric part of the
specific entropy. The SGS flux does not contribute to the net radial
energy transport because it is decoupled from the mean stratification,
and therefore changing $\chiSGS$ does not lead to drastic changes in
the boundary layer thickness near the surface.

The simulations were made using the {\sc Pencil
  Code}\footnote{\href{https://github.com/pencil-code/}{https://github.com/pencil-code/}}
\citep{2021JOSS....6.2807P}. In the present study the code employs
third-order temporal and sixth-order spatial
discretisation. Advective terms in \Equsa{equ:indu}{equ:ss} are
written as fifth-order upwinding derivatives with a sixth-order
hyperdiffusive correction where the diffusion coefficient is
flow-dependent; see Appendix~B of \cite{DSB06}.

\subsection{System parameters and diagnostics quantities}
\label{sec:syspar}

The simulations are defined by the energy flux imposed at the bottom
boundary, $\Fbot=-(K \pd T/\pd r)|_{r=r_{\rm in}}$, the values of
$K_0$, $a$, $b$, $\rho_0$, $T_0$, $\Omega_0$, $\nu$, $\eta$,
$\chiSGS$, the profile of $K$, and the value of the modified
Stefan-Boltzmann constant $\sigma_{SB}$ in the upper boundary
condition $\sigma_{\rm SB} T_{\rm surf}^4 = K\pd T/\pd r$, where
$T_{\rm surf}$ is the (unconstrained) surface temperature. The current
models use a significantly enhanced luminosity in comparison to real
stars to bring the thermal and dynamical timescales close enough to be
resolved in the simulations. This leads to correspondingly higher
convective velocities and therefore the rotation rate is increased
accordingly to capture a similar rotational influence on the flow in
the simulations in comparison to real stars; see appendix~A of
\cite{2020GApFD.114....8K}.

The non-dimensional luminosity is given by
\begin{eqnarray}
\mathcal{L}\,=\,\frac{L_0}{\rho_0 (GM)^{3/2}R^{1/2}},
\end{eqnarray}
where $\rho_0$ is the initial density at the base of the convection
zone. The degree of luminosity enhancement is given by the ratio
$L_{\rm ratio}=\mathcal{L}/\mathcal{L}_\odot \approx 2.1\cdot 10^5$,
where $\mathcal{L}_\odot$ is the dimensionless solar luminosity. The
initial stratification is determined by the non-dimensional pressure
scale height at the surface
\begin{eqnarray}
\xi_0\,=\,\frac{\mathcal{R}T_1}{GM/R},
\end{eqnarray}
where $T_1=T(R,t=0)$.

The relative strengths of viscosity, SGS diffusion, and magnetic
diffusivity are given by the SGS and magnetic Prandtl numbers
\begin{eqnarray}
\PraSGS = \frac{\nu}{\chiSGS}, \ \ \Pm =\frac{\nu}{\eta}.
\end{eqnarray}
We use $\Pm=1$ in most of the runs and vary $\PraSGS$ between 0.1 and
10. The thermal Prandtl number related to the radiative conductivity
is given by
\begin{eqnarray}
\Pr = \frac{\nu}{\chi},
\end{eqnarray}
where $\chi=K/\cP\rho$ is the radiative diffusivity, which in general
varies as a function of radius, latitude, and time. In the current
simulations $\chiSGS \gg \chi$ almost everywhere. The efficiency of
convection is quantified by the Rayleigh number
\begin{eqnarray}
  \Ra \,= \,\frac{GM (\Delta r)^4}{\nu \chi_{\rm r} R^2}\left(- \frac{1}{c_{\rm P}} \frac{{\rm d}s_{\rm hs}}{{\rm d}r}\right)_{r_{\rm s}},
\end{eqnarray}
where $\Delta r=0.3R$ is the depth of the layer, $s_{\rm hs}$ is the
specific entropy in a one-dimensional non-convecting hydrostatic
model, evaluated near the top of the domain at $r_{\rm s}=0.95R$, and
where $\chi_{\rm s}$ is the total thermal diffusivity $K/\cP\rho$ from
$r=r_{\rm s}$. The hydrostatic solution is Schwarzschild-unstable only
in a thin layer near the surface \citep[see, e.g.][]{BB14,Br16} which
is why the Rayleigh number is evaluated at $r_{\rm s}$. Moreover,
$\chiSGS$ does not contribute to $\Ra$ because it only acts on
deviations from the spherically symmetric specific
entropy. Additionally, a turbulent Rayleigh number is quoted:
\begin{eqnarray}
\Ra_{\rm t} \,= \,\frac{GM (\Delta r)^4}{\nu \chi_{\rm tot} R^2}\left(- \frac{1}{c_{\rm P}} \frac{{\rm d}\brac{s}_{\theta\phi}}{{\rm d}r}\right)_{r_{\rm s}},
\end{eqnarray}
where $\brac{s}_{\theta\phi}$ is the time- and horizontal average of
the specific entropy and $\chi_{\rm tot} = \chiSGS +
\brac{\chi}_{\theta\phi}$ is the total thermal diffusivity. $\Ra_{\rm
  t}$ is always significantly smaller than $\Ra$ because $\chiSGS \gg
\chi$.

The magnitude of rotation is controlled by the Taylor number
\begin{eqnarray}
  \Ta = \frac{4\Omega_0^2 (\Delta r)^4}{\nu^2}.
\end{eqnarray}
The fluid and magnetic Reynolds numbers and the P\'eclet number are
given by
\begin{eqnarray}
\Rey = \frac{\urms}{\nu k_1}, \ \  \ReM = \frac{\urms}{\eta k_1}, \ \ \Pe = \frac{\urms}{\chiSGS k_1},
\label{equ:Rey}
\end{eqnarray}
respectively, where $\urms = \sqrt{\threehalfs (U_r^2+U_\theta^2)}$ is
the time- and volume averaged rms velocity where $U_\phi^2$ has been
replaced by $(U_r^2+ U_\theta^2)/2$ to avoid contributions from
differential rotation. The inverse of the wavenumber $k_1=2\pi/\Delta
r\approx21/R_\odot$ is used to characterize the radial extent of the
convection zone. Several definitions of the Coriolis number that
describes the rotational influence on the flow are discussed in
\Seca{sec:DR}.

Mean quantities are denoted by overbars are defined by the time- and
azimuthal average:
\begin{eqnarray}
  \mean{f}(r,\theta) = \frac{1}{\Delta \phi \Delta t} \int_{t_0}^{t_0
    + \Delta t}\!\!\int_0^{\Delta \phi} f(r,\theta,\phi,t) d\phi dt,
\end{eqnarray}
where $t_0$ and $\Delta t$ are the beginning and the length of the
statistically steady part of the simulation, and where $\Delta \phi =
\pi/2$ is the azimuthal extent of the simulation domain. Error
estimates are obtained by dividing the time series in three parts and
computing averages over each one of them. The largest deviation of
these sub-averages from the average over the whole time series is
taken to represent the error.

\begin{table*}[t!]
\centering
\caption[]{Summary of the runs.}
  \label{tab:runs}
      $$
          \begin{array}{p{0.08\linewidth}ccccccccccc}
          \hline
          \hline
          \noalign{\smallskip}
          Run  & \PraSGS  & \Ta [10^6] & \Co  & \Co_\ell  & \Co_\omega  & \Co_\star  & \Rey  & \Pe  & \ReM  & \Ra_{\rm t} [10^6]  & {\rm  DR} \\
          \hline
          P01-1H  &  0.1  & 2.6 &  1.14  &  1.10  &  0.59  & 0.38 &  36  &  3.6  &   -   &  0.17  & \mbox{AS}   \\
          P01-2H  &  0.1  & 3.1 &  1.25  &  1.15  &  0.63  & 0.42 &  36  &  3.6  &   -   &  0.19  & \mbox{AS}   \\
          P01-3H  &  0.1  & 3.7 &  1.34  &  1.21  &  0.66  & 0.46 &  37  &  3.7  &   -   &  0.21  & \mbox{(AS)}   \\
          P01-4H  &  0.1  & 4.4 &  1.42  &  1.22  &  0.67  & 0.50 &  37  &  3.7  &   -   &  0.23  & \mbox{(AS)}   \\
          P01-5H  &  0.1  & 5.1 &  1.50  &  1.28  &  0.70  & 0.54 &  38  &  3.8  &   -   &  0.25  & \mbox{(SL)}   \\
          P01-6H  &  0.1  & 5.8 &  1.65  &  1.36  &  0.75  & 0.58 &  37  &  3.7  &   -   &  0.27  & \mbox{SL}   \\
          \hline
          P01-1M  &  0.1  & 2.6 &  1.21  &  1.12  &  0.62  & 0.38 &  34  &  3.4  &   34  &  0.19  & \mbox{AS}   \\
          P01-1Mh &  0.1  &  29 &  1.25  &  0.85  &  0.50  & 0.38 & 109  &   11  &  109  &   2.2  & \mbox{AS}   \\
          P01-2M  &  0.1  & 3.1 &  1.29  &  1.17  &  0.65  & 0.42 &  35  &  3.5  &   35  &  0.20  & \mbox{AS}   \\
          P01-2Mh &  0.1  &  35 &  1.33  &  0.74  &  0.40  & 0.42 & 113  &   11  &  113  &   2.4  & \mbox{(SL)}   \\
          P01-3M  &  0.1  & 3.7 &  1.35  &  1.20  &  0.66  & 0.46 &  36  &  3.6  &   36  &  0.21  & \mbox{AS}   \\
          P01-3Mh &  0.1  &  41 &  1.49  &  0.80  &  0.43  & 0.46 & 110  &   11  &  110  &   2.6  & \mbox{SL}   \\
          P01-4M  &  0.1  & 4.4 &  1.48  &  1.24  &  0.70  & 0.50 &  36  &  3.6  &   60  &  0.24  & \mbox{(SL)}   \\
          P01-5M  &  0.1  & 5.1 &  1.59  &  1.33  &  0.75  & 0.54 &  36  &  3.6  &   51  &  0.26  & \mbox{SL}   \\
          P01-6M  &  0.1  & 5.8 &  1.92  &  1.54  &  0.84  & 0.58 &  32  &  3.2  &   53  &  0.28  & \mbox{SL}   \\
          \hline
          P1-1H   &  1.0  & 2.6 &  1.31  &  1.00  &  0.55  & 0.38 &  31  &   31  &   -   &   1.4  & \mbox{AS}   \\
          P1-2H   &  1.0  & 3.1 &  1.48  &  1.13  &  0.61  & 0.42 &  30  &   30  &   -   &   1.5  & \mbox{AS}   \\
          P1-3H   &  1.0  & 3.7 &  1.63  &  1.15  &  0.64  & 0.46 &  30  &   30  &   -   &   1.7  & \mbox{AS}   \\
          P1-4H   &  1.0  & 4.4 &  1.80  &  1.21  &  0.68  & 0.50 &  30  &   30  &   -   &   1.8  & \mbox{AS}   \\
          P1-5H   &  1.0  & 5.1 &  1.97  &  1.25  &  0.73  & 0.54 &  29  &   29  &   -   &   2.0  & \mbox{SL}   \\
          P1-6H   &  1.0  & 5.8 &  2.14  &  1.32  &  0.78  & 0.58 &  29  &   29  &   -   &   2.1  & \mbox{SL}   \\
          \hline
          P1-1M   &  1.0  & 2.6 &  1.42  &  0.98  &  0.58  & 0.38 &  29  &   28  &   29  &   1.6  & \mbox{AS}   \\
          P1-2M   &  1.0  & 3.1 &  1.58  &  1.08  &  0.63  & 0.42 &  28  &   28  &   28  &   1.7  & \mbox{AS}   \\
          P1-2Mh  &  1.0  &  35 &  1.56  &  0.68  &  0.38  & 0.42 &  96  &   96  &   96  &    20  & \mbox{(SL)}   \\
          P1-3M   &  1.0  & 3.7 &  1.72  &  1.13  &  0.67  & 0.46 &  29  &   29  &   29  &   1.8  & \mbox{AS}   \\
          P1-3Mh  &  1.0  &  41 &  1.73  &  0.74  &  0.41  & 0.46 &  94  &   94  &   94  &    21  & \mbox{SL}   \\
          P1-4M   &  1.0  & 4.4 &  1.87  &  1.20  &  0.71  & 0.50 &  28  &   28  &   28  &   1.9  & \mbox{SL}   \\
          P1-5M   &  1.0  & 5.1 &  2.03  &  1.26  &  0.76  & 0.54 &  28  &   28  &   28  &   2.0  & \mbox{SL}   \\
          P1-6M   &  1.0  & 5.8 &  2.19  &  1.33  &  0.80  & 0.58 &  28  &   28  &   28  &   2.1  & \mbox{SL}   \\
          \hline
          P10-1M  &   10  & 2.6 &  1.75  &  0.99  &  0.58  & 0.38 &  23  &  233  &   23  &    13  & \mbox{AS}   \\
          P10-2M  &   10  & 3.1 &  1.98  &  1.06  &  0.64  & 0.42 &  23  &  227  &   23  &    14  & \mbox{AS}   \\
          P10-3M  &   10  & 3.7 &  2.16  &  1.16  &  0.70  & 0.46 &  23  &  227  &   23  &    15  & \mbox{AS}   \\
          P10-4M  &   10  & 4.4 &  2.31  &  1.26  &  0.74  & 0.50 &  23  &  229  &   23  &    15  & \mbox{AS}   \\
          P10-5M  &   10  & 5.1 &  2.53  &  1.31  &  0.80  & 0.54 &  23  &  226  &   23  &    16  & \mbox{(AS)}   \\
          P10-6M  &   10  & 5.8 &  2.68  &  1.40  &  0.84  & 0.58 &  23  &  228  &   23  &    17  & \mbox{(SL)}   \\
          P10-7M  &   10  & 6.6 &  2.80  &  1.46  &  0.88  & 0.62 &  23  &  234  &   23  &    17  & \mbox{(AS)}   \\
          P10-8M  &   10  & 7.5 &  3.10  &  1.56  &  0.95  & 0.65 &  22  &  224  &   22  &    18  & \mbox{SL}   \\
          P10-9M  &   10  & 8.4 &  3.28  &  1.62  &  1.00  & 0.69 &  22  &  224  &   22  &    18  & \mbox{SL}   \\
          \hline
          \end{array}
          $$ \tablefoot{Hydrodynamic (MHD) runs are denoted by suffix
            H (M). Majority of the runs uses grid resolution
            $128\times288\times144$, the exception being runs denoted
            by the suffix h where resolution $256\times576\times288$
            was used. The Rayleigh number in most runs is $7.0 \cdot
            10^8$, expect in the higher resolution runs, denoted by
            suffix h, where it is $\Ra = 2.4\cdot10^9$. $\Co$,
            $\Co_\ell$, $\Co_\omega$, and $\Co_\star$ denote the
            definitions of the Coriolis number given by
            Eqs.~(\ref{equ:Cori}), (\ref{equ:Col}), (\ref{equ:Cow}),
            and (\ref{equ:Costar}). $\PrM=1$ in all runs except in
            P01-4M and P01-6M where $\PrM=1.67$, and in P01-5M where
            $\PrM=1.43$ to obtain growing dynamos. The last column
            denotes the type of differential rotation with parenthesis
            indicating that the result is not statistically
            significant.}
\end{table*}

\begin{figure*}
  \begin{center}
  \includegraphics[width=.9\textwidth]{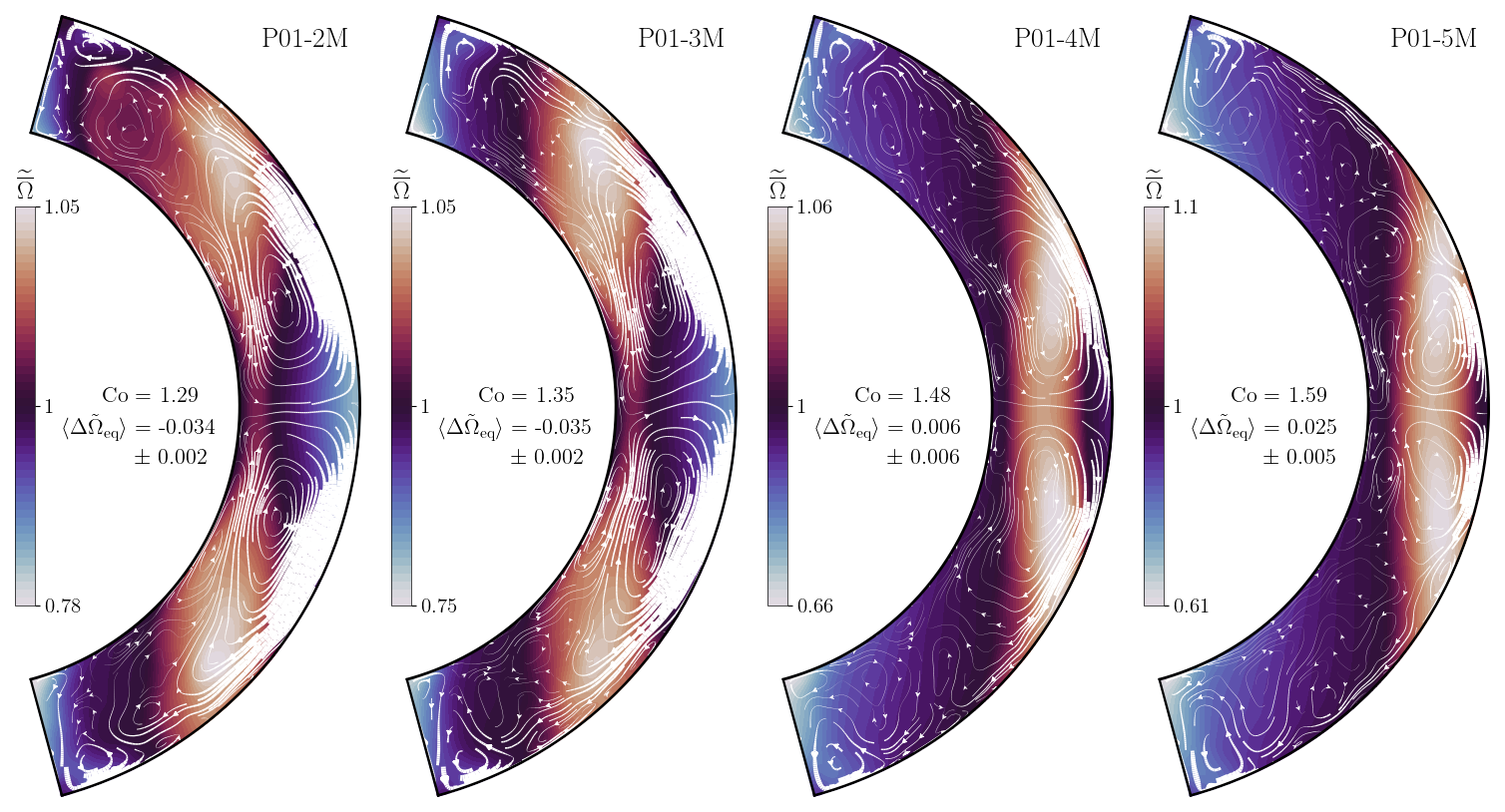}
  \end{center}
  \caption{Time-averaged rotation profiles from MHD runs P01-[2-5]M
    near the AS-SL transition with $\PraSGS = 0.1$ with rotation rate
    increasing from left to right. Coriolis numbers according to
    \Eq{equ:Cori} and the mean differential rotation at the equator
    $\brac{\Delta \tilde{\Omega}_{\rm eq}}$ according to
    \Eq{equ:DOmega_eq} are indicated in each panel. The white arrows
    indicate the meridional flow.}
\label{fig:pOm2_m_Pr01}
\end{figure*}

\subsection{Initial and boundary conditions}

Initially the stratification is isentropic with polytropic index
$n=1.5$ and $\xi_0=0.02$, resulting in an initial density contrast of
30. The value of $K_0$ is chosen such that $F_{\rm rad}=F_{\rm tot}$
at the bottom of the domain.

The radial and latitudinal boundaries are assumed impenetrable and
stress-free for the flow. On the bottom boundary, a fixed heat flux is
prescribed while at the top a black body condition is applied. On the
latitudinal boundaries, the gradients of thermodynamic quantities are
set to zero; see \cite{KMCWB13}. For the magnetic field we apply a
radial field condition at the upper, and a perfect conductor condition
at the lower boundary. On the latitudinal boundaries the field is
assumed to be tangential to the boundary. These conditions are given
in terms of the magnetic vector potential by:
\begin{eqnarray}
A_r & = & 0,\ \ \ \frac{\pd A_\theta}{\pd r} = -\frac{A_\theta}{r},\ \ \ \frac{\pd A_\phi}{\pd r} = -\frac{A_\phi}{r}\ \ \ (r=R),\\
\frac{\pd A_r}{\pd r} & = & A_\theta = A_\phi = 0\ \ \ (r=\rin), \\
A_r & = & \frac{\pd A_\theta}{\pd \theta} = A_\phi = 0\ \ \ (\theta=\theta_0,\pi-\theta_0).
\end{eqnarray}
The azimuthal direction is periodic for all quantities. The velocity
and magnetic fields are initialized with random low-amplitude Gaussian
noise fluctuations.

\section{Results} \label{sect:results}

Three sets of simulations were done where $\PraSGS = 0.1$ (set P01),
$1$ (P1), and $10$ (P10), respectively. The first two sets contain
hydrodynamic and MHD runs, and a subset of the MHD runs were remeshed
to higher resolution and correspondingly higher Rayleigh, P\'eclet,
and Reynolds numbers; see \Table{tab:runs}. Only MHD variants of the
P10 runs were run.

\begin{figure*}
  \begin{center}
  \includegraphics[width=.9\textwidth]{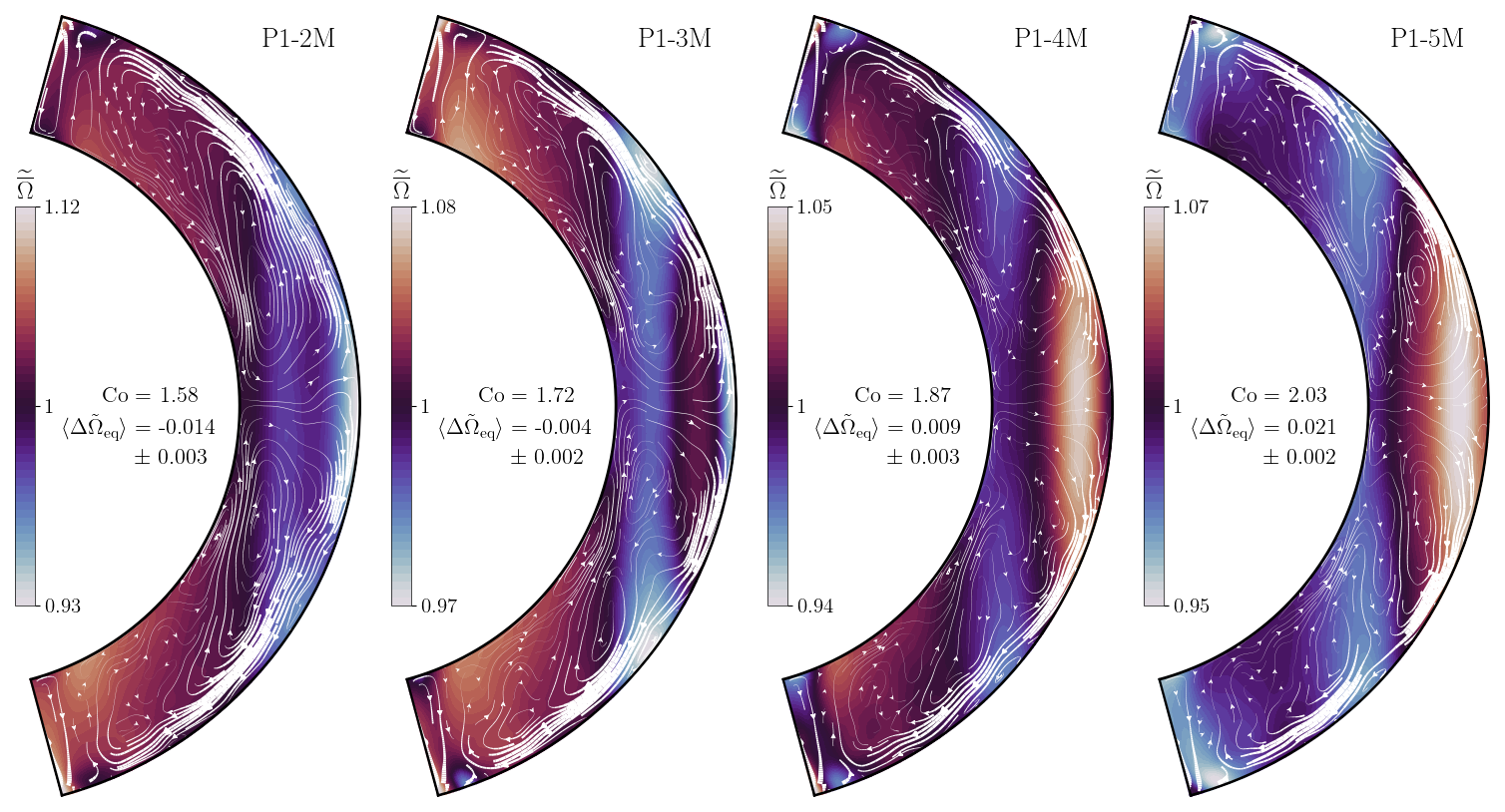}
  \end{center}
  \caption{Same as \Figa{fig:pOm2_m_Pr01} but for runs P1-[2-5]M in set
    P1.}
\label{fig:pOm2_m_Pr1}
\end{figure*}

\begin{figure*}
  \begin{center}
  \includegraphics[width=.9\textwidth]{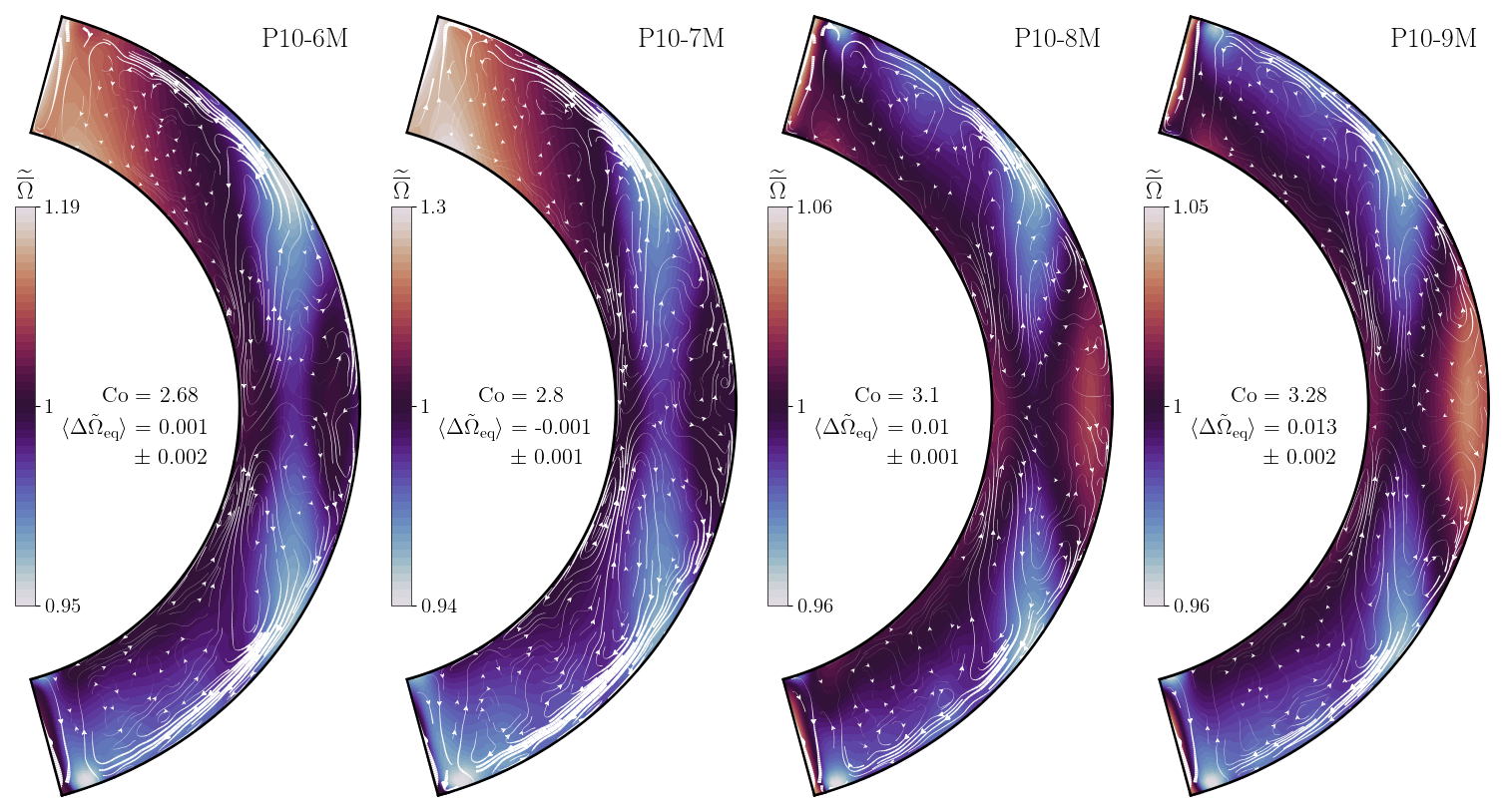}
  \end{center}
  \caption{Same as \Figa{fig:pOm2_m_Pr01} but for runs P10-[6-9]M in
    set P10.}
\label{fig:pOm2_m_Pr10}
\end{figure*}

\begin{figure*}
  \begin{center}
    \includegraphics[width=\columnwidth]{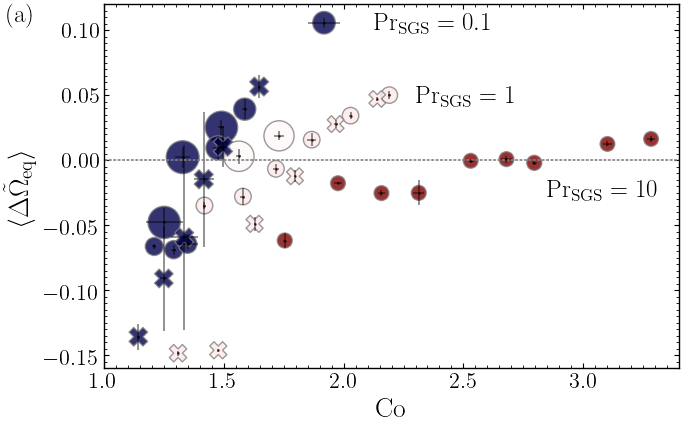}\includegraphics[width=\columnwidth]{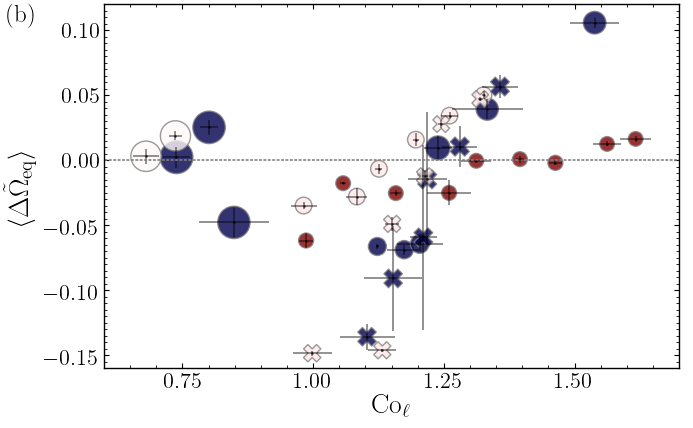}
    \includegraphics[width=\columnwidth]{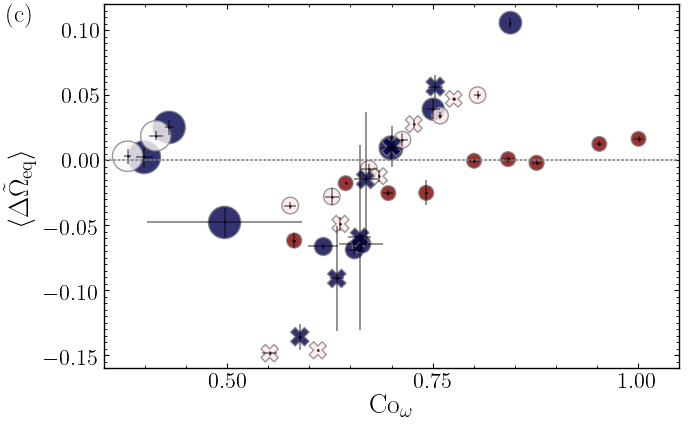}\includegraphics[width=\columnwidth]{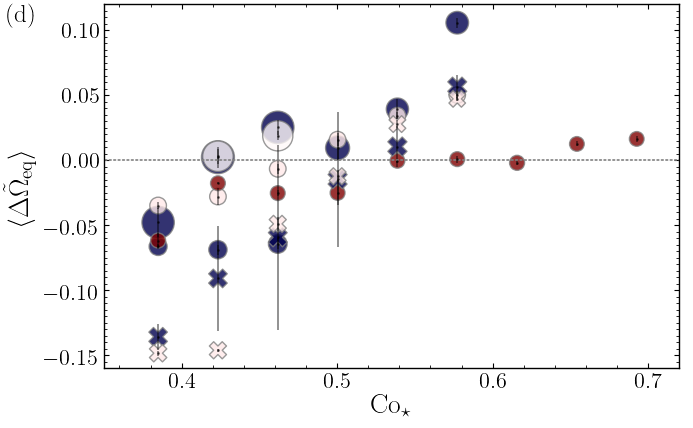}
  \end{center}
  \caption{Normalized average angular velocity at the equator
    $\brac{\Delta \tilde{\Omega}_{\rm eq}}$ for all runs as a function
    of $\Co$ {\it (a)}, $\Co_\ell$ {\it (b)}, $\Co_\omega$ {\it (c)},
    and $\Co_\star$ {\it (d)}. Circles (crosses) denote MHD
    (hydrodynamic) runs and the colour of the symbols indicates the
    SGS Prandtl number. The sizes of the symbols are proportional to
    $\ReM$ ($\Rey$) for MHD (hydrodynamic) runs.}
\label{fig:pOmega_eqt}
\end{figure*}

\subsection{Differential rotation and meridional circulation}
\label{sec:DR}

The main focus of the current study is to explore the effects of the
Prandtl number for the large-scale flows that develop in rotating
convective systems. The mean rotation profile is given by time- and
azimuthal average:
\begin{eqnarray}
  \mean{\Omega}(r,\theta) = \Omega_0 + \frac{\mUUp(r,\theta)}{r \sin \theta},
\end{eqnarray}
and the meridional flow is given by $\mUUU_{\rm mer} =
(\mUUr,\mUUt,0)$. In many simulations the latitudinal profiles of
$\mean{\Omega}$ are non-monotonic such that the rotation rate has a
polar jet, a maximum at mid-latitudes or sometimes several local
minima and maxima as function of latitude. Furthermore, equatorial
asymmetries can occur, rendering the amplitude of the latitudinal
shear an unreliable diagnostic of the overall sense of differential
rotation; see representative examples in
\Figust{fig:pOm2_m_Pr01}{fig:pOm2_m_Pr10}. Therefore the
classification of AS and SL rotation profile is here based on the mean
rotation profile at the equator
\begin{eqnarray}
\brac{\Delta \tilde{\Omega}_{\rm eq}} = \frac{\int_{r_{\rm in}}^{R} r^2 [\tilde{\mOm}(r,\theta_{\rm eq}) - 1] dr}{\int_{r_{\rm in}}^{R} r^2 dr},\label{equ:DOmega_eq}
\end{eqnarray}
where $\theta_{\rm eq} = \pi/2$, and where the tildes refer to
normalization by the rotation rate of the frame of reference,
$\Omega_0$. If $\brac{\Delta\tilde{\Omega}_{\rm eq}} > 0$
($\brac{\Delta\tilde{\Omega}_{\rm eq}} < 0$) the run is classified as
SL (AS) rotator. This measure turns out to be a monotonic function of
rotation and it is furthermore unaffected by equatorial asymmetries or
latitudinal jets.

The current results indicate that the convective velocity increases
when the Prandtl number is decreased. This is manifested by increasing
fluid Reynolds number for decreasing SGS Prandtl number; see the
eighth column of \Table{tab:runs}. Naively one could then expect that
achieving SL differential rotation for low $\PraSGS$ would be more
difficult, that is, require faster rotation. Often the rotational
influence on the flow is quantified by a simple definition of the
Coriolis number
\begin{eqnarray}
\Co\,=\,\frac{2\,\Omega_0}{\urms k_1},
\label{equ:Cori}
\end{eqnarray}
where the convective length scale is assumed to be unchanged by
rotation. Using this definition to characterize the results, the
Coriolis number at which the rotation profile flips from AS to SL
appears to decrease monotonically as $\PraSGS$ decreases; see
\Figa{fig:pOmega_eqt}(a) where $\brac{\Delta \tilde{\Omega}_{\rm eq}}$
is shown for all runs as a function of $\Co$. That is, in the low
resolution MHD runs with $\PraSGS=0.1$ ($1$) the transition occurs
around $\Co\approx1.5$ ($\Co\approx1.8$); see
\Figus{fig:pOm2_m_Pr01}{fig:pOm2_m_Pr1}, whereas for $\PraSGS=10$ the
transition occurs at an even higher Coriolis number ($\Co\approx3$);
see \Figa{fig:pOm2_m_Pr10}.

However, the validity of this simplistic definition of the Coriolis
number to characterize the simulations can be questioned based on its
very crude estimate of the convective length scale and velocity
amplitude. For example, \cite{GYMRW14} argued that a local Rossby
(inverse Coriolis) number based on the length scale from the mean
spherical harmonic degree $\mean{\ell}_u$ of the $m\neq0$ poloidal
flows gives a more accurate estimate \citep[see
  also][]{SPD12}. Furthermore, they showed that the scatter near the
AS-SL transition is reduced when using this definition, essentially
reducing the apparent dependence on Prandtl number significantly. Here
this was tested by computing $\mean{\ell}_u$ according to
\begin{eqnarray}
\mean{\ell}_u = \sum_\ell \ell \frac{\brac{\UUU_{\rm p}^\ell \bm\cdot \UUU_{\rm p}^\ell}}{\brac{\UUU_{\rm p} \bm\cdot \UUU_{\rm p}}},
\end{eqnarray}
where $\UUU_{\rm p}$ is the non-axisymmetric poloidal flow and the
superscript $\ell$ refers to the corresponding spherical harmonic
degree. Data from a varying number of horizontal slices from near the
base, at the middle, and near the top of the CZ were analyzed for each
run, and the resulting $\mean{\ell}_u$ is an average over the depths
and time. The number of time slices per run varies between 7 and
roughly 60. The corresponding length scale is $d_u = \pi \Delta
R/\mean{\ell}_u$, and the Coriolis number based on this is given by
\begin{eqnarray}
\Co_\ell = \frac{2\Omega_0 d_u}{\urms},
\label{equ:Col}
\end{eqnarray}
where $\urms$ is defined the same way as in \Eq{equ:Rey}. The numbers
in \Table{tab:runs}, fifth column, indicate that the value of
$\Co_\ell$ is sensitive to the Reynolds number such that in the runs
with the highest $\Rey$ the values of $\Co_\ell$ are roughly 30 per
cent smaller than those of the low resolution runs. Similarly, a
Coriolis number based on fluctuating vorticity can be defined as
\citep[e.g.][]{2022ApJ...926...21B},
\begin{eqnarray}
  \Co_\omega = 2\Omega_0/\orms',
\label{equ:Cow}
\end{eqnarray}
where $\ooo' = \bm\nabla\times\uuu$ with $\uuu = \UUU - \mUUU$. This
quantity shows a similar sensitivity to the fluid Reynolds number as
$\Co_\ell$; see the sixth column of \Table{tab:runs}. Both of these
definitions pick up smaller length scales at the highest Reynolds
number runs resulting in lower Coriolis numbers. This is likely an
indication that the simulations are still far away from an asymptotic
regime where the results would be independent of the
diffusivities. Nevertheless, $\Co_\ell$ and $\Co_\omega$ characterize
the rotational influence on the flow more accurately than $\Co$ by
being sensitive to the actual dominant length scale. However, due to
the Reynolds number sensitivity, one should only compare results from
runs with comparable $\Rey$ when these definitions are used to
characterize the results.

The average radial differential rotation at the equator is shown as
functions of $\Co_\ell$ and $\Co_\omega$ in
Figures~\ref{fig:pOmega_eqt}(b) and (c), respectively. Ignoring the
five runs at higher Reynolds and P\'eclet numbers for the time being,
these results suggest that the dependence of the differential rotation
transition as a function of the Prandtl number all but vanishes for
$\PraSGS < 1$. However, in both cases the transition for $\PraSGS =10$
occurs still at a higher $\Co_\ell$ and $\Co_\omega$ than for the
$\PraSGS =0.1$ and $1$ cases\footnote{Note that only the two most
  rapidly rotating runs with $\PraSGS=10$ have statistically
  significant SL differential rotation; see the 12th column of
  \Table{tab:runs}.}. These results are in accordance with those
reported by \cite{2018PhFl...30d6602K} who also found that a Prandtl
number above unity promotes AS differential rotation due to enhanced
downward angular momentum transport. Both definitions give much lower
Coriolis numbers for the higher-$\Rey$ runs because smaller convective
scales are resolved and $\mean{\ell}_u$ and $\omega'$ pick these up.
Furthermore, among the five higher-$\Rey$ runs, the single AS model
(P01-1Mh) appears to have a marginally larger Coriolis number than the
SL counterparts in both cases, although the horizontal error estimates
are large in both cases. Whether this is a real effect or due to
insufficient statistics remains open at this point. As a side-note, the
resemblance of Figures~\ref{fig:pOmega_eqt}(b) and (c), or
alternatively the correlation between $\Co_\omega$ and $\Co_\ell$,
suggests that $\Co_\omega$ captures the rotation dependence of the
convective length scale almost as well as $\Co_\ell$ without having to
perform numerically expensive spherical harmonic decomposition.

Each of the definitions of the Coriolis number discussed so far rely
on diagnostic quantities such as $\urms$, $\mean{\ell}_u$, and
$\omega'$ that are sensitive to other details, such as the fluid
Reynolds number, of the system. Yet another alternative is to define a
Coriolis number that depends only on stellar input parameters such as
the luminosity, mass, and rotation rate of the modeled star. This can
be constructed by assuming that
\begin{equation}
L = \rho_\star u_\star^3 R^2,
\end{equation}
where $L=4\pi \rin^2 \Fbot$ is the luminosity, $\rho_\star$ is a
reference density, and $u_\star$ is an estimate of an average
convective velocity. Here we assume $\rho_\star =
\rho_0$\footnote{Using the average density of the star, $\rho_{\rm av}
  = M/{4 \over 3} \pi R^3$, is another option and in that case the
  definition of $\Co_\star$ is fully determined by stellar
  parameters.}  and construct a stellar Coriolis number
\begin{equation}
\Co_\star = \frac{2 \Omega_0 R}{u_\star} = \frac{2 \Omega_0 R^{5/3}\rho_0^{1/3}}{L^{1/3}}.
\label{equ:Costar}
\end{equation}
While this definition is imperfect in the sense that the actual
convective flow speed or scale do not enter, it is useful in
determining whether a model with a given rotation rate and luminosity
is an AS or SL rotator. This makes particularly sense in the
homogeneous set of simulations considered here where the stellar mass,
luminosity, and radius are all fixed. The results are shown in
\Figa{fig:pOmega_eqt}(d). It is apparent that the runs with
$\PraSGS=10$ require a significantly higher $\Co_\star$, corresponding
here to higher $\Omega_0$, to achieve SL differential
rotation. However, the AS-SL transition occurs at the same $\Co_\star$
for $\PraSGS=1$ and $0.1$. Notably the MHD runs at $\PraSGS=0.1$ and
$1$ for $\Co_\star=0.50$ are at least marginally SL in comparison to
the corresponding AS hydrodynamic cases. Similarly the higher
resolution runs at $\Co_\star=0.46$ are SL or marginally SL for
$\Co_\star=0.46$, whereas the corresponding lower resolution MHD and
hydrodynamic runs are all AS. The current higher resolution runs with
$\ReM\approx 94\ldots 113$ also have small-scale dynamos which was
tested with separate runs where the axisymmetric ($m=0$) magnetic
fields were suppressed. Therefore it is plausible that the main
contribution to the earlier appearance of SL differential rotation in
these cases is due to the growing importance of the magnetic fields.
However, no corresponding higher resolution hydrodynamic runs were
made to confirm this. The current results are in accordance with the
results of \cite{2021NatAs...5.1100H} and \cite{2022arXiv220204183H}
who argue in favour of the magnetic field being the decisive factor in
the transition. Finally, if the simulations are scaled to physical
units as in \cite{2020GApFD.114....8K}, the lowest value in the
present study, $\Co_\star = 0.38$, corresponds to the case of solar
rotation rate at solar luminosity.

\begin{figure*}
    \includegraphics[width=\textwidth]{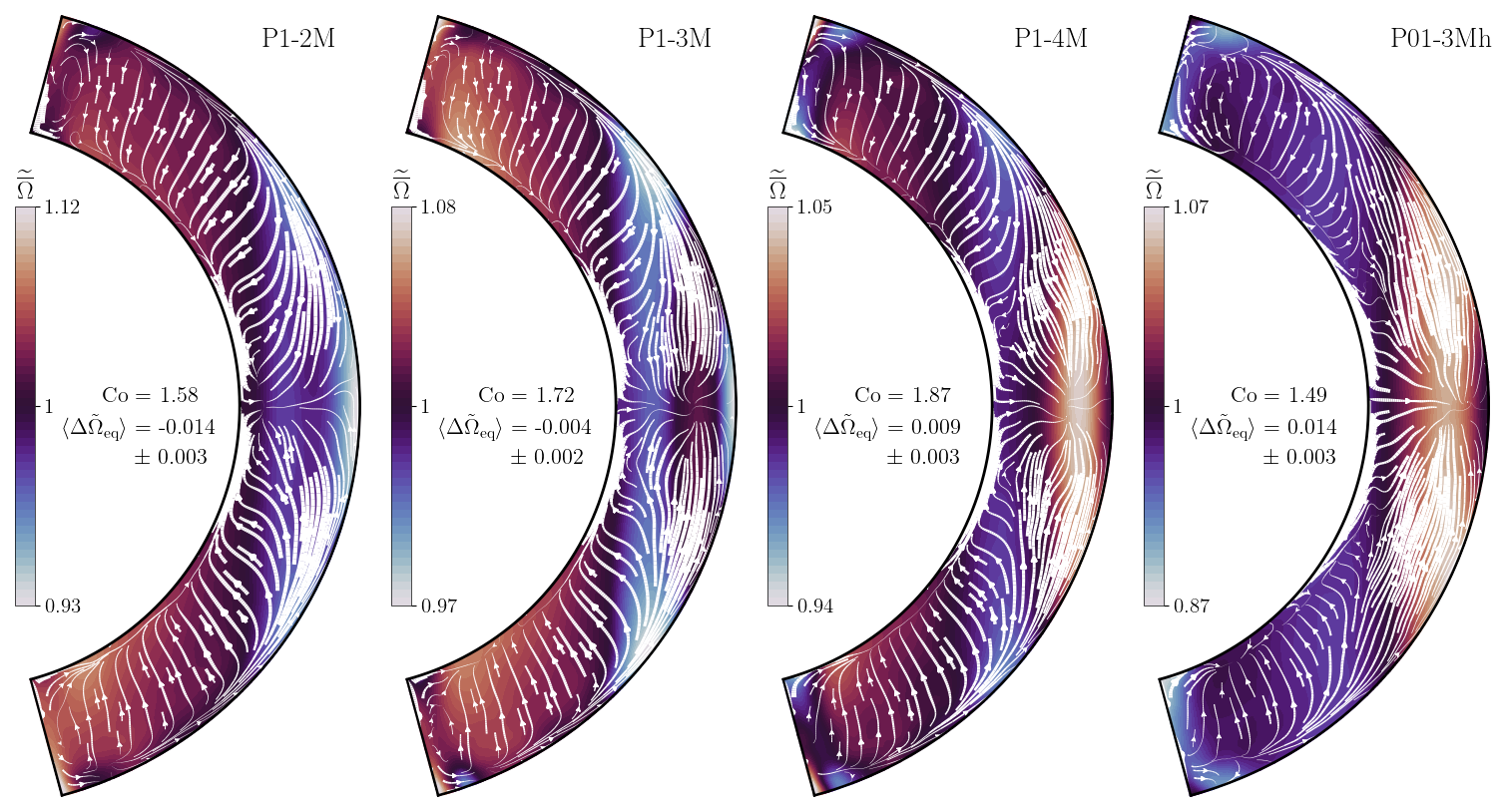}
  \caption{Total angular momentum flux $\bm{\mathcal{T}}$ (arrows)
    superimposed on the mean angular velocity profile (colour
    contours) from runs P1-[2-4]M and P01-3Mh.}
\label{fig:pvec_stress}
\end{figure*}

The conclusion of comparing the results characterized in terms of the
four variants of the Coriolis number is that SL differential rotation
is substantially more difficult to obtain for $\PraSGS = 10$ than for
$\PraSGS = 1$ and $0.1$, whereas in the latter two cases there is no
clear difference. Furthermore, the use of the simplistic Coriolis
number, \Eq{equ:Cori}, gives misleading results and should be
avoided. Contrary to the SGS Prandtl number, the dependence on
magnetic fields is clearer such that in MHD runs SL differential
rotation is easier to excite.

\begin{figure*}
  \begin{center}
    \includegraphics[width=.95\textwidth]{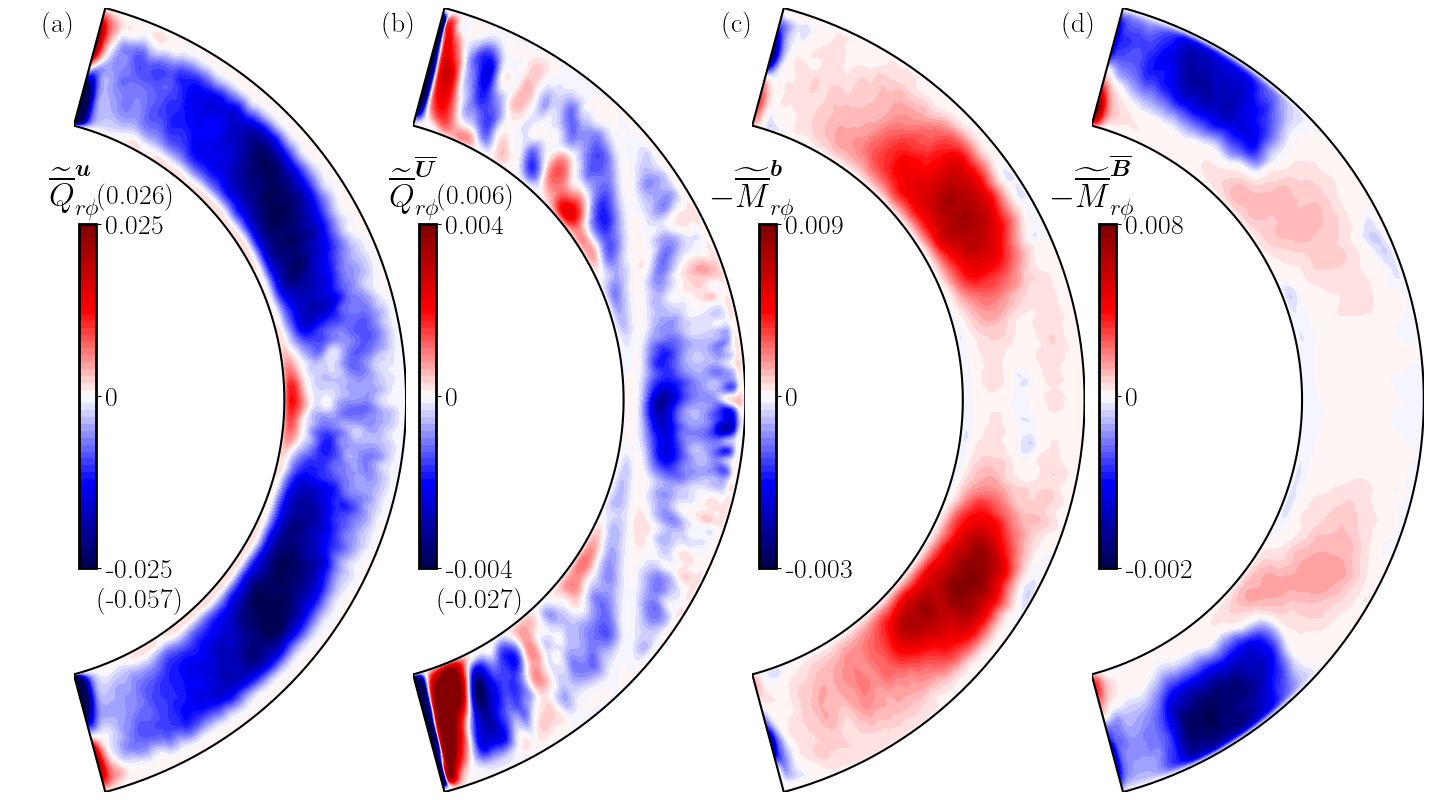}
    \includegraphics[width=.95\textwidth]{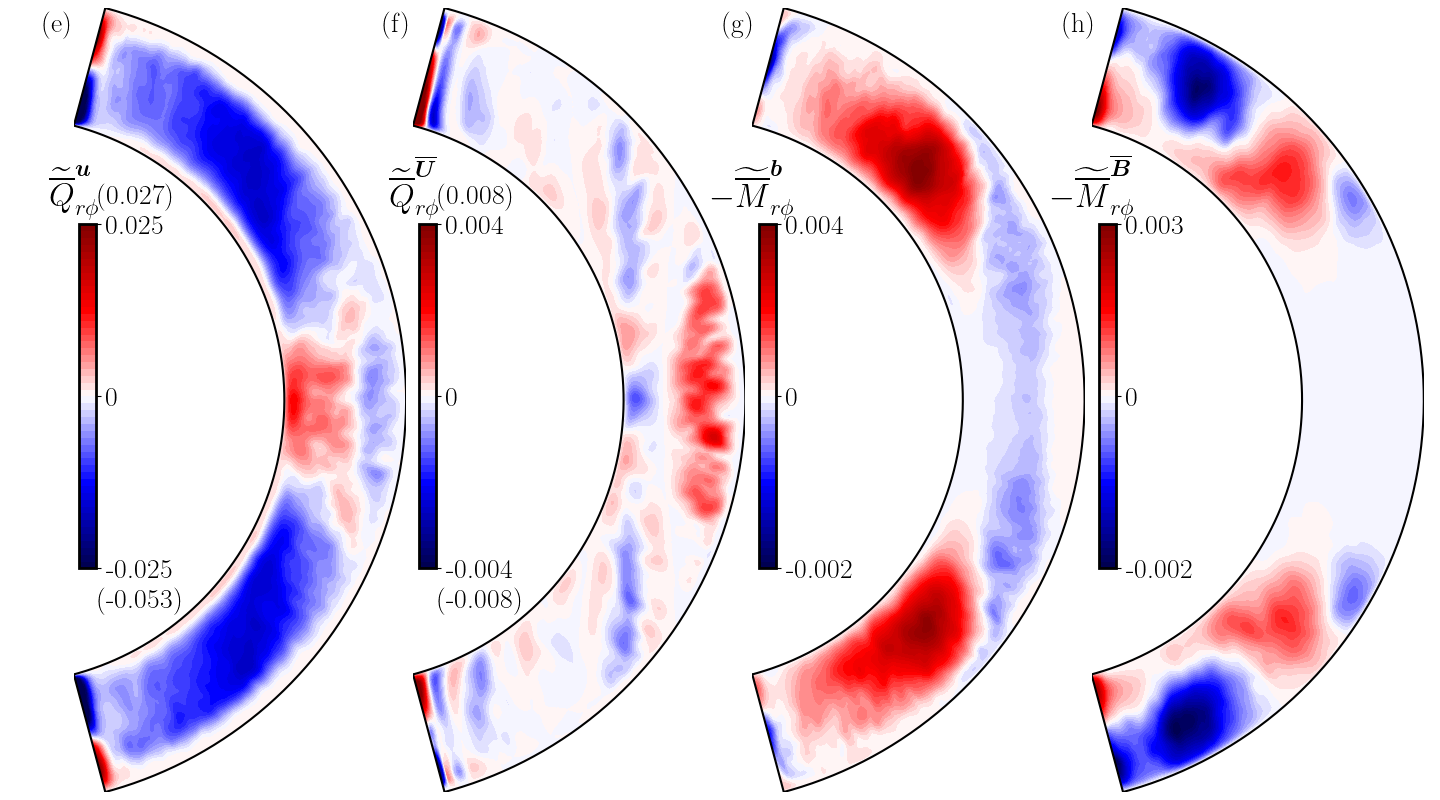}
  \end{center}
  \caption{Time-averaged contributions to the radial angular momentum
    flux for runs P1-2M (top row; panels {\it a} to {\it d}) and P1-4M
    (bottom; panels {\it (e)} to {\it (h)}). The tildes refer to
    normalization by $\rho_0 \urms^2$.}
\label{fig:prad_stress}
\end{figure*}

\subsection{Angular momentum transport}

When contributions from molecular viscosity can be neglected, the
angular momentum in the interior of the star is governed by the
conservation equation
\begin{equation}
\frac{\pd}{\pd t} (\mean{\rho}{\mathcal L}) + \bm\nabla\bm\cdot (r \sin\theta \bm{\mathcal{T}}) = 0,
\end{equation}
where ${\mathcal L} = r^2 \sin^2\theta \mOm$ is the specific angular
momentum and
\begin{eqnarray}
\bm{\mathcal{T}} = \mean{\rho}(\mean{\uuu u_\phi} + \mUUU\ \mUUp) - (\mean{\bbb b_\phi}+\mBBB\ \mBBp)/\mu_0,
\end{eqnarray}
is the total flux of angular momentum. The internal angular velocity
profile is thus determined by the spatial distribution of the angular
momentum fluxes. The main transporters are due to the Reynolds and
Maxwell stress due to fluctuating and mean flows and fields
\citep[cf.][]{R89,RH04}
\begin{eqnarray}
\mean{Q}_{ij}^\uuu &=& \mean{\rho}\mean{u_i u_j}, \\
\mean{M}_{ij}^\bbb &=& -\mean{b_i b_j}/\mu_0, \\
\mean{Q}_{ij}^{\mUUU} &=& \mean{\rho}\mUUi\ \mUUj, \\
\mean{M}_{ij}^{\mBBB} &=& -\mBBi\ \mBBj/\mu_0,
\end{eqnarray}
where $\bbb = \BBB - \mBBB$ is the fluctuating magnetic field.

Representative results of $\bm{\mathcal{T}}$ are shown in
\Figa{fig:pvec_stress} superimposed on top of the angular velocity in
Runs~P1-[2-4]M and P01-3Mh. The total angular momentum flux is nearly
radially downward at high latitudes in the bulk of the CZ. The region
of downward radial flux is roughly confined inside the tangent
cylinder in all cases. Outside the tangent cylinder and near the
surface at all latitudes, $\bm{\mathcal{T}}$ is directed predominantly
equatorward irrespective whether the rotation profile is AS or
SL. Outside the tangent cylinder the angular momentum transport is
increasingly axial although the radial flux at the equator remains
non-zero in all cases. The sense of the differential rotation is
related to the sign of the radial component of $\bm{\mathcal{T}}$ at
the equator: for positive (outward) flux the differential rotation is
SL (Runs~P1-4M and P01-3Mh in \Figa{fig:pvec_stress}) whereas it is AS
for negative (downward) flux (Run~P1-2M). In the transitory case of
Run~P1-3M the flux at the equator converges at the local maxima of
$\mOm$ similarly as in the SL cases of P1-4M and P01-3Mh. Therefore it
appears sufficient to study the radial angular momentum flux to
determine the difference between AS and SL differential rotation.

\Figu{fig:prad_stress} shows all of the components of the radial
angular momentum flux for runs P1-2M (P1-4M) with AS (SL) differential
rotation. In both cases the Reynolds stress due to $m \neq 0$
contributions of the velocity is the dominant contributors while the
Reynolds stress due to meridional circulation and Maxwell stresses are
clearly subdominant. Moreover, in the AS case P1-2M the radial
transport due to the Reynolds stresses is downward almost everywhere,
while in the SL case P1-4M, both $\mean{Q}_{r\phi}^\uuu$ and
$\mean{Q}_{r\phi}^{\mUUU}$ are positive near the equator. The sign of
the latter near the equator is determined by the sign of $\mUUp$
because $\mUUr > 0$ in all cases; see the meridional flow, for
example, in \Figa{fig:pOm2_m_Pr01}.

\begin{figure}
  \begin{center}
    \includegraphics[width=\columnwidth]{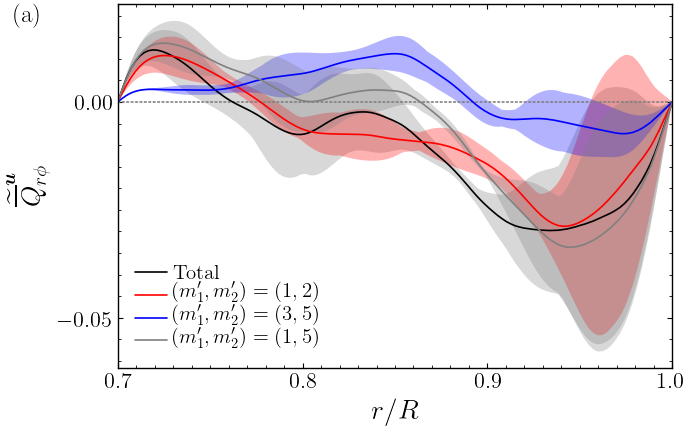}
    \includegraphics[width=\columnwidth]{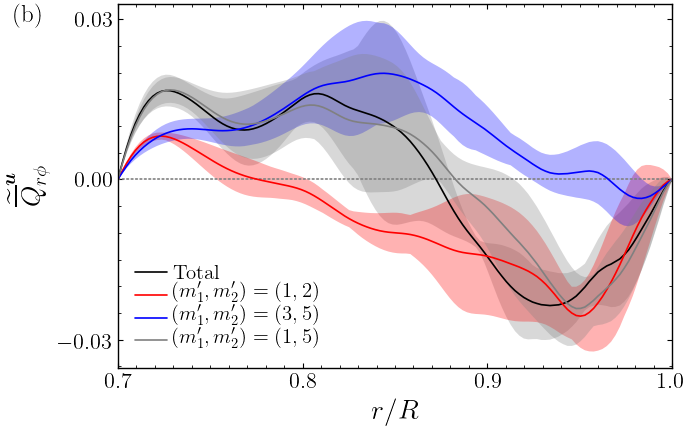}
  \end{center}
  \caption{Fourier-filtered and total Reynolds stress component
    $\widetilde{\mean{Q}}_{r\phi}^\uuu$ from runs P1-2M {\it (a)} and
    P1-4M {\it (b)} as indicated in the legend. The shaded areas
    indicate error estimates according to the definition in
    \Seca{sec:syspar}.}
\label{fig:Rxz_m}
\end{figure}

\begin{figure}[t]
  \begin{center}
    \includegraphics[width=\columnwidth]{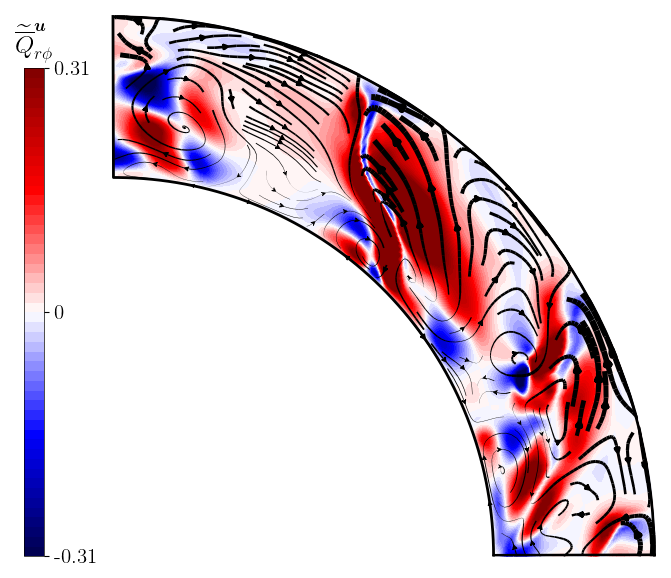}
  \end{center}
  \caption{Instantaneous normalized radial Reynolds stress component
    $\widetilde{\mean{Q}}_{r\phi}^\uuu$ on the equatorial plane from
    run P01-6M (colour contours). The $m\neq0$ flows are indicated by
    the arrows, the width of which is proportional to the local flow
    amplitude.}
\label{fig:pBusse_mOm1.5_Pr01_s3}
\end{figure}

To study the equatorial Reynolds stress in more detail, a number of
slices of the $m\neq0$ velocity field $\UUU(r,\theta_{\rm eq},\phi)$
were analyzed. Azimuthal Fourier filtering was applied to produce
filtered velocity fields $\UUU_{\rm f}^{(m'_1,m'_2)}$, where azimuthal
orders ranging from $m'_1$ to $m'_2$ were retained\footnote{Note that
  due to the $\Delta\phi = \pi/2$ azimuthal extent of the simulation
  domain, $m'$ corresponds to $4m$ in a full sphere.}. These flows are
used to compute the Reynolds stress
\begin{equation}
\mean{Q}_{r\phi}^{(m'_1,m'_2)}\!=\!\frac{1}{\Delta\phi \Delta t} \int_{t_0}^{t_0+\Delta t}\!\!\!\int_0^{\Delta \phi}\!\!\!\rho U_r^{(m'_1,m'_2)} U_\phi^{(m'_1, m'_2)} d\phi dt.\label{equ:Qff}
\end{equation}
Here the density fluctuations are assumed to be small and no Fourier
filtering was applied to $\rho$. Representative results are shown in
\Figa{fig:Rxz_m}. In the AS case P1-2M, \Figa{fig:Rxz_m}(a), the total
Reynolds stress is negative everywhere except at the very base of the
CZ. Contributions from the largest scale $(m'_1,m'_2 = 1,2)$
non-axisymmetric motions are statistically almost identical with the
total stress. A weak positive contribution around the middle of the CZ
is visible for $(m'_1,m'_2 = 3,5)$ but the Reynolds stress for
$(m'_1,m'_2 = 1,5)$ is again very similar to the total stress. In the
SL runs the largest scales $(m'_1,m'_2 = 1,2)$ also contribute to a
downward flux whereas the main contribution to the net outward flux
comes from $(m'_1,m'_2 = 3,5)$; see \Figa{fig:Rxz_m}(b) for
Run~P1-4M. Similarly to the AS case, the contributions from $m'>5$ are
very small. This indicates that practically all of the Reynolds stress
at the equator is due to relatively large-scale structures which can
be identified as the Busse columns
\citep{1970ApJ...159..629B,1970JFM....44..441B} which are also often
referred to as banana cells. Such features are often prominently
visible in snapshots of the velocity field; see
\Figa{fig:pBusse_mOm1.5_Pr01_s3} for a representative example. The
Busse cells are manifestations of non-linear prograde-propagating
thermal Rossby waves. It is therefore somewhat questionable to talk
about turbulent Reynolds stress in this context since the Busse
columns are large-scale convective modes that appear essentially at a
scale corresponding to forcing of turbulence.

The mechanism by which the differential rotation is generated in the
current simulations is therefore different from that in
\cite{2022arXiv220204183H} and \cite{2022arXiv220204183H} where the
small-scale Maxwell stress that is the dominant contribution to the
radial angular momentum transport. While the high-resolution runs in
the present study have small-scale dynamos and show an increased
tendency for SL differential rotation, the Reynolds stress due to the
thermal Rossby waves is still the dominant contribution to the angular
momentum flux in all of the runs considered here. In an earlier study
\citep{2017A&A...599A...4K} the Maxwell stresses were found to be
comparable to the Reynolds stress at the highest magnetic Reynolds
numbers, but in that study the modeled stars were rotating typically
three to four times faster than in the current study. Although the
Maxwell stress dominates the angular momentum transport in the
simulations of \cite{2022arXiv220204183H}, large-scale Busse columns
can still be seen in the deep parts of their model; see, for example,
their Fig.~6. If such large-scale convective patters were as prominent
in the Sun, they should have been detected by helioseismology but
there is no evidence currently to this effect. Therefore it seems that
although highly magnetized simulations are more solar-like in terms of
the rotation profile, the conundrum with the too prominent large-scale
structures remains.

Furthermore, in hydrodynamic mean-field theories of differential
rotation \citep[e.g.][]{R89,KR05,2015JPlPh..81e3904R} the turbulence
models are necessarily simplified and the large-scale convection modes
such as the Busse columns do not appear. In the most commonly adopted
approach of \cite{KR05}, the radial angular momentum transport is
downward for slow, and vanishing for rapid rotation at the
equator. The SL differential rotation results in from a strong
equatorward transport. Numerical simulations of isothermal homogeneous
anisotropic turbulence also produce downward (slow rotation) or
vanishing (rapid rotation) radial angular momentum flux at the equator
\citep{Kap19} in qualitative accordance with \cite{KR05}. Hydrodynamic
mean-fields models based on these concepts do not typically produce AS
solutions unless strong magnetic fields are present
\citep{2004AN....325..496K}, although more recently a hydrodynamic
mechanism has also been discussed \citep{RKKS19}. This latter process
relies on poleward horizontal angular momentum flux at slow rotation
which was also found from local simulations, but which appears to be
absent in global models such as those presented here. The mean-field
theories avoid the problem of too prominent too strong thermal Rossby
waves by simply neglecting them, whereas they fail to characterize
both AS and SL cases in the current simulations. This tension is yet
another facet of the convective conundrum, the resolution of which is
likely to require further critical assessment of both theoretical and
simulation approaches.

\section{Conclusions}

The transition from AS to SL differential rotation was studied as a
function of the SGS Prandtl number ($\PraSGS$). Four definitions of
the Coriolis number were used to quantify the exact point of
transition from simulations where the rotation of the star was
varied. While this transition occurs at a higher Coriolis number for
$\PraSGS=10$ than for $\PraSGS=1$ and $0.1$, no statistically relevant
difference was found between the last two cases. This suggests that
the Prandtl number dependence of the AS-SL transition is weak for
$\PraSGS < 1$, whereas a high Prandtl number makes it significantly
more difficult to achieve SL differential rotation.

These results are puzzling because earlier non-rotating local
simulations \citep{2021A&A...655A..78K} suggested that also the cases
$\Pra=1$ and $\Pra=0.1$ differ significantly in many respects. A
notable difference to the study of \cite{2021A&A...655A..78K} is that
the current simulations do not include a radiative layer below the
CZ. This can explain why no subadiabatic layers develop at the base of
the CZs in the current simulations because the effects of overshooting
are absent. The latter was found to be particularly sensitive to the
Prandtl number in the local simulations \cite{2019A&A...631A.122K} and
\cite{2021A&A...655A..78K}. The inclusion of a radiative layer has
also consequences for the dynamo solutions
\citep[e.g.][]{GSdGDPKM15,2022ApJ...931L..17K} which also couple back
to differential rotation. These aspects need to be revisited in future
studies.

Many of the current simulations also included magnetic fields, albeit
often in a parameter regime where the small-scale dynamo is not
excited. Thus the influence of magnetic fields is relatively weak in
most of the current runs. Nevertheless, the magnetic fields make it
easier to excite SL differential rotation especially in the current
higher resolution runs that likely also have small-scale
dynamos. However, the effects of magnetic fields are likely to be more
significant in more realistic higher-$\ReM$ systems as manifested by
the recent results of \cite{2022arXiv220204183H}. Therefore magnetism
appears to be the most promising candidate to explain the discrepancy
between solar observations and global simulations. Nevertheless, the
non-detection of thermal Rossby waves, which are still prominent in
all current simulations, from the Sun still raises questions as to the
generation mechanism of solar differential rotation. Finally, the
tension between mean-field theories of differential rotation and 3D
simulation results is pointed out as another aspect that requires
further scrutiny in the future.

\begin{acknowledgements}
  I acknowledge the hospitality of Nordita during the program `The
  Shifting Paradigm of Stellar Convection: From Mixing Length Concepts
  to Realistic Turbulence Modelling'. The simulations were made within
  the Gauss Center for Supercomputing project ``Cracking the
  Convective Conundrum'' in the Leibniz Supercomputing Centre's
  SuperMUC--NG supercomputer in Garching, Germany. This work was
  supported by the Deutsche Forschungsgemeinschaft Heisenberg
  programme (grant No.\ KA 4825/4-1).
\end{acknowledgements}

\bibliographystyle{aa}
\bibliography{paper}

\end{document}